# What is the simplest model which captures the basic experimental facts of the physics of underdoped cuprates?

H. Alloul

Laboratoire de Physique des Solides, UMR CNRS 8502, Université Paris-Sud 11, 91405, Orsay (France)

**Abstract.** The discovery of cuprates has been underlined by two salient phenomena, their high temperature superconductivity (SC) and the occurrence of the "pseudogap" (PG) in the underdoped part of the hole doped phase diagram, with a large set of associated fascinating physical properties. A large debate has been opened more than ten years ago about the actual significance of the pseudogap: signature of preformed pairs or independent order in competition with SC. This debate was apparently on the verge to be cleared in favour of the competing order scenario. Indeed a series of recent experiments done with distinct experimental techniques report charge ordered or symmetry broken states in the underdoped cuprates. Some speculations based on theoretical calculations suggest that those observations reveal a totally different origin than expected initially for the superconductivity in the cuprates. These new speculations justify recalling some of the experimental results which have been established on the early days of the HTSC and which have been quite often "forgotten", or "overlooked" by newcomers in the field who tend to consider that the pseudogap onset temperature $T^*$ is an- ill defined temperature. I shall recall here that there is no doubt whatsoever from the early days that the onset of the pseudogap state is much higher in temperature than all the recently detected charge order phenomena in comparable samples, which apparently compete with SC. (Indeed, the onsets of these new states are detected at temperatures at which a 50% decrease of the spin susceptibility has already occurred) I shall also recall that some of the early mean field theories have initiated the idea that both SC and the PG could be generated within a simple doped Hubbard model, or its simplified t-J model version. Recent calculations extending those to Dynamical Mean Field Theories (DMFT), with more or less sophisticated versions give theoretical determinations of physical quantities which begin to resemble quite nicely most of the experimental observations done below $T^*$. Direct comparisons are still not secured as they do rely on the approximations done in the DMFT calculations. The basic Hubbard model does not generate the CDW phases, presumably due to symmetry breaking perturbations present in the actual materials. Though we had underlined in the past that the disorder of the chemical dopants is one of the dominant perturbations, the recent experiments on chain ordered YBCO suggest that the ordering of the dopants can indeed be influential as well in stabilizing commensurate CDW states. Therefore, the variety of broken symmetry states detected recently at lower temperatures than the pseudogap $T^*$ certainly require more complicated modelling and are not necessarily generic. They most certainly correspond to specific ground states of the pseudogap electronic matter which exhibit interesting correlated orders.

The French Science Academy honoured my scientific activity in 2012 by awarding me a great prize " Science and Innovation " subsidized by the CEA, for establishing in Orsay an NMR research group dedicated to the physics of correlated electron systems. I have indeed been involved initially up to the 1980s on the physics of Kondo impurities in metals, of spin glasses, or on the metal insulator transitions in Phosphorus doped silicon, which were my first contacts with correlated electron physics. But, although those achievements have acquired international recognition, the experimental activities done with my students and colleagues in Orsay which are most frequently cited and acknowledged in the correlated electron international community are certainly those linked with the high $T_c$ cuprate superconductors. Though we have been quite active in that field since the discovery of the cuprates and have done several seminal discoveries, those were initially only poorly recognized within the French national community. I do consider that this prize represents indeed the first national recognition of our contributions in that field.

I have been consequently asked to write an article summarizing the work which justified the attribution of the prize, and I therefore decided to focus here on the physics of the cuprates and specifically on the occurrence of the pseudogap phase in the phase diagram which I contributed to reveal from the early days of the high $T_c$ physics. During those early experiments we immediately revealed some unique properties of the weakly doped cuprates and the occurrence of the pseudogap which only became a few years later a key feature in the phase diagram of the cuprates, when it could be investigated by various experimental techniques. Since that, many new experimental observations of specific phenomena on the pseudogap phase have oriented the study of the superconductivity in cuprates in novel directions and many speculations on the origin of the pseudogap have been done. I have therefore maintained a continuing interest on those proposals and on the various experimental observations and performed some specific experiments which permitted us to remain extremely active in that field.

The physics of cuprates remains therefore a leading challenge in the correlated electron physics which has been constantly renewed by the development of fascinating new technologies which reveal unexpected experimental behaviours. Although those have been mostly done on the experimental side, some recent developments of numerical calculations using DMFT techniques have also triggered my interest. At the time at which this prize was awarded to me by the Science Academy, I was writing a viewpoint on these recent developments which I posted on the arXiv and submitted successively for publication in many journals. The fact that this viewpoint has not been published so far (*) is a direct proof that the field remains extremely controversial and that many points raised are worth further investigations. I therefore reproduce below the text of this viewpoint, the abstract given above being the only part modified with respect to the original version.

## 1 Introduction

The discovery of cuprates has been underlined by two salient phenomena, their high temperature superconductivity (SC) and the occurrence of the "pseudogap" (PG) in the underdoped part of the hole doped phase diagram. The latter was revealed first by the observation of a sharp decrease of the NMR Knight shift $K$ [1] that is of the spin susceptibility $\chi_s$ of the $CuO_2$ plane. This has been followed by the discovery of associated features in the specific heat $C_v$ [2], in the in plane $\rho_{ab}$ and out of plane $\rho_c$ resistivities [3] and in the optical properties [4]. In the last decades fascinating physical effects have been highlighted in the PG regime, such as the persistence of SC signatures above $T_c$ [5], **k** space differentiation detected in the normal state by ARPES [6] and revealed by the scattering on intrinsic disorder in STM experiments [7]. However the transition at $T^*$ to the PG regime has been considered as a crossover, as no sharp feature has been evidenced so far on the thermodynamic properties. I shall comment first recent theoretical developments in comparison with these experiments, and shall discuss later the various broken symmetry sates detected recently within the PG regime.

## 2 Mean field and C-DMFT solutions of the Hubbard model

The proximity to a Mott insulator at half filling has always been considered as a key point in the physics of the doped cuprates [8]. It has been anticipated that a purely electronic mechanism could induce a $d$-wave SC state, which was established experimentally more than fifteen years ago [9] [10]. Within a Hubbard model on a square lattice, or its simplified $t$-$J$ model version, it has been shown rather early-on that a mean field solution ("slave bosons") could yield a pseudogap in the underdoped regime, which would disappear and merge with $d$-wave SC in the overdoped regime [8].

However these early mean field solutions seemed to imply a scenario in which preformed pairs occur above $T_c$ and in which the PG line would merge with the SC line upon overdoping. This approach did not match well with the early experimental observations which indicated that the PG line drops fast towards the SC dome [1]. Furthermore, the PG being robust to disorder, contrary to SC, allowed us initially to suggest that these were distinct phenomena [11]. These points have been supported by detailed analyses [12] of the variation of $T^*$ versus doping, which will be recalled later hereafter.

Recent DMFT calculations permit to extend the solutions of the Hubbard model beyond simple mean field theory. Those have to be performed beyond the single site approach [13], in order to reproduce the momentum space anisotropy seen in ARPES [14]. Large cells might be required in these cellular (C-DMFT) approaches to get a full description of the Fermi surface properties [15]. However, in a series of papers, Sordi *et al* do evidence that calculations with a simple two by two plaquette suffice to recover the differentiation between SC and the pseudogap [16]. Solving the C-DMFT equation with Quantum Monte Carlo methods, they could perform systematic computations of the $T$ dependence of the thermodynamic and transport properties, and proposed a C- DMFT derived phase diagram for the cuprates in which the first order transition from a PG phase to a correlated metal with increasing doping is disrupted by the occurrence of $d$-wave SC. This transition evolves above a critical point into a PG crossover, which delineates the regime in which strong correlations and Mott physics dominate. This so called Widom line bears a close resemblance with the experimental $T^*$ line.

One might wonder then whether these computations are reaching a stage allowing quantitative comparisons with experiments. There are reasons leading one to be cautious in that respect, which I'll discuss later. But a valuable outcome of these C-DMFT calculations that I will underline is that they permit one to compare the signatures of the pseudogap on the different physical properties. This will allow me to confirm that $T^*$ is not an ill defined temperature, contrary to the perception which has been induced by the large set of low $T$ phenomena discovered recently in the PG regime.

## 3 Experimental and C-DMFT determinations of $T^*$

Let us remark first that the decrease of the *Knight shift $K_s$ or spin susceptibility* is an unambiguously large effect corroborated in many cuprate families. Indeed this quantity drops progressively on a wide $T$ range from its large normal state nearly $T$ independent value to zero in the SC state (Fig. 1). Such a behaviour of $K_s(T)$ is reproduced quite well in the computations by Sordi *et al*.

If the electronic transformation reflected by the NMR shift would monitor a second order phase transition, one would consider the onset $T^*$ of the $K_s$ decrease as the transition temperature at which the order parameter sets in. But, the transition being rather a smooth crossover, one could be tempted to define the PG from the low $T$ spectral width $\Delta$ of the pseudogap. As initially done by Loram *et al* [2], in an analysis of the *NMR and specific heat data*, one could assume that a $T$ independent PG density of states (DOS) is thermally populated. This allowed them to define the energy scale $\Delta$ for the pseudogap, which depends of the chosen shape for the DOS (triangular or $d$-wave) [12].

Remarkably, the C-DMFT results being numerical data, Sordi *et al* faced exactly the same problems as for the experiments in characterizing the PG state.

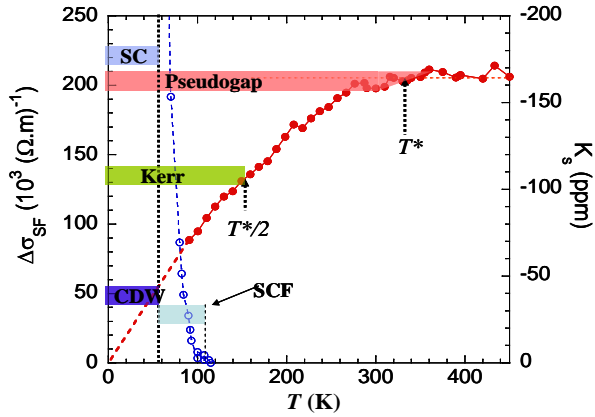

Fig 1 Spin contribution $K_s$ to the $^{89}$Y NMR Knight shift [11] for YBCO$_{6.6}$ permit to define the PG onset $T^*$. Here $K_s$ is reduced by a factor two at $T \sim T^*/2$. The sharp drop of the SC fluctuation conductivity (SCF) is illustrated (left scale) [23]. We report as well the range over which a Kerr signal is detected [28], and that for which a CDW is evidenced in high fields from NMR quadrupole effects [33] and ultrasound velocity data [30]. (See text).

They initially estimated it [17] at the maximum of $d\chi/dT$, which would corresponds to an average gap $\Delta$. Then in their recent paper [16] they find that one could as well use the maximum of $\chi$ to define the onset $T^* \sim 2\Delta$.

The *in plane transport* $\rho_{ab}(T)$ has also been found early on to display a characteristic S shape variation for underdoped samples [3] [18] [19]. This permits one to define $\Delta$ at the inflection point and $T^*$ at the onset of the high $T$ linear variation (see Fig 2). Detailed experimental comparisons [18] established that $[\rho_{ab}(T)- \rho_{ab}(0)]/T$ and $K_s(T)$ have quite identical $T$ variations, leading to $\Delta \sim T^*/2$.

Sordi *et al* did not compute $\rho_{ab}(T)$, but focussed on calculations of the *out of plane transport* $\rho_c(T)$, which switches from metallic to semiconducting behaviour in the PG regime. They showed that the minimum of $\rho_c(T)$ relates perfectly with $T^*$ deduced from $K_s(T)$. Although it is harder to take reliable data for $\rho_c(T)$, one systematic study [20] on BiSCO films permitted to define a similar $T^*$ onset as that obtained from $\rho_{ab}(T)$.

Let me recall as well that $\Delta$ and $T^*$ are not affected even by large in plane disorder, as has been emphasized initially on NMR shift data for Zn substituted YBCO [11]. Similarly Mathiessen's rule is perfectly obeyed by $\rho_{ab}(T)$, at least at high $T$, as could be monitored (Fig 2) on pure single crystals in which defects have been introduced by high energy electron irradiation [21]. This insensitivity to disorder is important as it reveals that $\Delta$ and $T^*$ are *generic features of the cuprates*, which permits comparisons between diverse samples and cuprates families.

Overall I would like to underline here that, in my view, an important agreement between the experiments and C-DMFT calculations is that the physical properties exhibit *smooth crossover variations*, which onset at a temperature $T^*$, such that the spin susceptibility decreases by about a factor two between $T^*$ and $\Delta \sim T^*/2$.

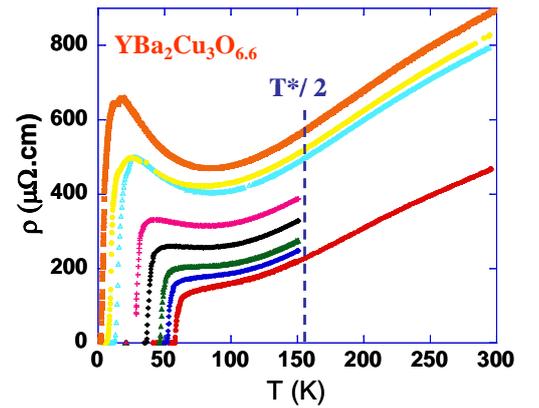

Fig 2 In plane transport data [21] in YBCO6.6 permit to define $\Delta \sim T^*/2$ from the inflexion point and $T^*$ from the onset of the deviation with respect to the high $T$ linear variation (which occurs above room $T$). Above 100K the S shape is independent of the concentration of defects induced by electron irradiation while $T_c$ decreases by more than a factor three (see text).

## 4  Widom line and PG line

Note however that quantitative comparisons between these CDMFT calculations and experiments are still not secured since a two by two cellular plaquette might not be sufficient to investigate the situation for which long range correlations develop at the lowest $T$ in the pseudogap regime. It has for instance been found that the AF correlation length can be as large as five unit cells in YBCO$_{6.6}$ both by NMR and neutron scattering [22].

But one issue that can still be pursued is the comparison of the phase diagrams, which has been a very debated issue in the community. The main question is whether one might decide the fate of the $\Delta$ or $T^*$ line with increasing doping, when the PG evolves towards the SC dome? In their joint analyses of the $T$ dependences of $K_s$, $\rho_{ab}$ and the specific heat, Tallon and Loram [12] could confirm that the value of $\Delta$ decreases abruptly towards the top of the SC dome, as was as well apparent from the onset $T^*$ values [1]. This has led them to suggest the occurrence of a critical point for slight overdoping.

However, when $T^*$ approaches $T_c$ one could notice that the occurrence of SC fluctuations (SCF) above $T_c$

gives deviations of the physical properties from the normal state that are similar to those caused by the onset of the pseudogap. So, most attempts to obtain experimentally the variation of $T^*$ versus doping were not reliable enough unless SC could be suppressed, for instance by disorder, or by an applied field. This has been done recently by applying large enough pulsed fields to suppress the SCF contributions to $\rho_{ab}(T)$, permitting us [23][24] to determine $T^*$ values just above $T_c$. These experiments gave evidence that the $T^*$ line hits the SC dome just above optimal doping, and that the onset of the SCF follows the SC dome and not the $T^*$ line, in clean samples with minimal disorder. So far, there are clear evidences from ARPES and STM that the PG spectral features persist in the SC state below $T^*$ in underdoped samples. However, there is no unambiguous experiment revealing a *PG onset below $T_c$ above optimal doping*.

While SC can hardly be avoided in the experiments, one can selectively privilege the normal state correlations or the SC ones in numerical C-DMFT techniques. So the Widom line could be found by Sordi *et al* to evolve into a first order transition which should occur at a doping $\delta_c$ in the absence of SC. The fact that no sign of the pseudogap remains for $\delta > \delta_c$ while SC persists, is therefore a very good qualitative agreement between C-DMFT results and the experimental situation. These C-DMFT computations permit then to remove the difficulties initially found in the mean field solutions of the Hubbard model.

The occurrence of SC has always been preventing determinations of the low $T$ thermodynamic properties in the PG regime. However, most of the extensive studies performed in the early days revealed the presence of a metallic-like component, similar to the one detected on the initial $^{89}$Y NMR data, which we had labelled as Fermi-liquid like [1], but would be considered nowadays as a bad metal. Similarly, the low $T$ transport can be fitted with a $T^2$ variation above $T_c$, and appears to continue to lower $T$ when $T_c$ is reduced by an applied field [25]. However the $T=0$ limit could usually not be checked as the PG state is unstable towards new metallic states as discussed below.

## 5 Broken symmetries

The activity on the cuprates has been dominated in this millennium by sets of new results obtained with advanced techniques, reporting broken symmetry states or ordered states occurring in the pseudogap regime.

The order detected by neutron scattering [26] has to be highlighted first since its onset occurs at temperatures which can reliably be identified with the $T^*$ values obtained by NMR shift and transport data. Furthermore the change of structure factor in the Bragg scattering has been detected in both YBCO and Hg1201 families and could be reliably assigned to a PG "phase" with some long range order. It had been anticipated for long by Varma [27], that one had to consider explicitly the oxygen in the $CuO_2$ planes unit cell to generate such an ordered state. This might explain the absence of identification of an order parameter in the Hubbard model C-DMFT. However, as the numerics reproduce most thermodynamic features, I am tempted to consider that the order introduced by the oxygen appears as a minor perturbation which stabilizes the ordered state without any essential incidence on the basic physical phenomena induced by the Mott correlations. On the experimental side, this would be corroborated by the fact that all features at $T^*$ are insensitive to disorder.

Most other new broken symmetries of PG matter have been most often considered as primary signatures of the PG regime by their discoverers. They are however found to occur at temperatures definitively lower than $T^*$. The onset of a Kerr effect signal [28] occurs in YBCO at a temperature $T_K \sim T^*/2$ for which the spin susceptibility has already been reduced by a factor two. The CDW order anticipated from transport [29] and quantum oscillations [30] is detected in high fields below $T_{CDW} \sim 50K$ [31]. So those can only be assigned to successive electronic structure transformations of the electronic PG matter occurring well below $T^*$, which could be specific to materials details. One important effort still required is to select among these physical phenomena those which are really generic of the cuprates.

We had underlined in the past the dominant incidence of the in plane disorder, or of that induced by chemical dopants [23]. The well known "stripe" phase studied at length in the LaBaCuO family [32], and the NMR evidence that the CDW order which occurs in YBCO samples in high applied fields is commensurate for chain ordered ortho II samples [33] imply that the order of chemical dopants does play a role in stabilizing these low $T$ phases.

This mirrors the situation that we established [34] in our extensive study of layered cobaltates $Na_xCoO_2$. There we have evidenced that for slightly different $x$ values the distinct Na orderings which occur correspond to quite different electronic ground states of the $CoO_2$ planes. For these cobaltates, the coupling to the Na lattice is large enough to stabilize pure Na ordered phases, while in YBCO the oxygen ordering energies appear smaller and do not permit the stabilization of perfect order.

To conclude, these C-DMFT calculations within the basic Hubbard model do generate most of the high $T$ specific properties of the pseudogap and SC in cuprates. Furthermore, these C-DMFT results suggest that the charge ordered phases evidenced experimentally result from instabilities of the PG matter to symmetry breaking perturbations which cannot be avoided in the actual materials. Those generate new fascinating states of correlated matter which often, but not always, compete with SC.

I would like to acknowledge F. Rullier-Albenque for many fruitful exchanges about the experimental work on transport properties. Together with G. Sordi and A. M. Tremblay, many theoretician colleagues such as M. Civelli, A. Georges, G. Kotliar and M. Rozenberg have helped me over the years to appreciate the new theoretical possibilities made available by C-DMFT calculations.